\documentclass[10pt]{article}
\usepackage{graphicx}
\usepackage{epsfig}
\usepackage{amsbsy}

\begin{document}

\makeatletter \Large \baselineskip 10mm \centerline{\bf Nonorthogonal tight-binding model for hydrocarbons}

\vskip 3mm

\centerline{M. M. Maslov, A. I. Podlivaev, L. A. Openov$^*$}

\vskip 3mm

\centerline{\it Moscow Engineering Physics Institute}
\centerline{\it (State University), 115409 Moscow, Russia}

\vskip 3mm

$^*$e-mail: LAOpenov@mephi.ru

\vskip 3mm

\centerline{\bf Abstract}

Parameters of the nonorthogonal tight-binding model for hydrocarbons are derived based on a criterion of the best agreement between the calculated and experimental values of bond lengths and binding energies for different molecules C$_{n}$H$_{m}$. The results obtained can be used, e. g., to study the kinetics of hydrogen absorption by carbon nanostructures, to simulate the dynamics of hydrocarbon clusters like cubane $\mathrm{C_{8}H_{8}}$, etc.

\vskip 20mm


\newpage

\centerline{\bf 1. Introduction}

In recent years, there was a growing interest in interaction between hydrogen (both atomic and molecular) and various carbon structures. This is in part due to the promising results of the first experiments on hydrogen absorption by carbon nanotubes \cite{Dillon,Liu,Ye} and a corresponding perspective to use hydrocarbons in hydrogen energetics, see reviews \cite{Eletskii,Nechaev}. A solid based on clusters $\mathrm{C_{8}H_{8}}$ (cubanes) \cite{Eaton} is one more example of hydrocarbon systems important from both fundamental and practical viewpoints. Cubane and its derivatives are considered as candidates to a new type of fuel, they can probably be used in pharmaceutics, liquid crystals, etc. \cite{Eaton2}.

In the case that experimental data remain incomplete and controversial, theoretical calculations become of particular importance. For simulations on hydrocarbons, either classical interatomic potentials, e. g., Tersoff-Brenner potential \cite{Tersoff,Brenner}, or so called first principle methods, e. g., the density functional theory (DFT) \cite{Hohenberg,Kohn}, are commonly employed. One should keep in mind, however, that making use of {\it ab initio} approaches do not guarantee, by itself, the reliable results since those approaches are sometimes extremely sensitive to the type of exchange-correlation potential, the set of the basis functions, the cutoff energy, etc. (for example, different modifications of DFT disagree on the most stable $\mathrm{C}_{20}$ isomer, see references in \cite{Sokolova}). Besides, the first principle calculations are computer-time consuming and put severe restrictions on the size of the system and/or the real-time interval during which the system evolution can be followed (usually $\sim 1 - 10$ ps).

In such a situation, tight-binding models are a reasonable compromise between more rigorous {\it ab initio} and oversimplified empirical calculations. Contrary to classical potentials, those models explicitly account for the contribution of electron subsystem to the total energy. Although the tight-binding method is not as strict as {\it ab initio}
approaches, it is competitive with them in the accuracy and not computer-resource-intensive, thus facilitating the simulation of relatively large systems and/or dynamic processes at real-time scales up to $\sim 1~ \mu$s. For carbon systems, various tight-binding models were suggested in Refs. \cite{Tomanek,Xu,Menon,Tang,Mehl}.
They were successfully used by many authors to simulate both carbon clusters and bulks. In particular, previously we made use of a model \cite{Xu} to study the thermal stability of fullerenes $\mathrm{C}_{20}$ and $\mathrm{C}_{60}$, one-dimensional chains and two-dimensional complexes of fullerenes $\mathrm{C}_{20}$, etc. \cite{Davydov,Podlivaev,Openov,Openov2,Podlivaev2,Openov3,Podlivaev3,Davydov2}.

Different tight-binding models for hydrocarbons were suggested in Refs. \cite {Davidson,Wang,Porezag,Horsfield,Winn,Pan,Zhao}. Each of those models has its own advantages and drawbacks, but all of them rather accurately describe a broad range of $\mathrm{C_{n}H_{m}}$ molecules and macroscopic C-H systems (for example, hydrogen on the diamond surface). Among the papers mentioned above, of particular interest is Ref. \cite{Zhao} whose authors generalized the tight-binding model to C-H-O systems containing oxygen along with carbon and hydrogen. The model \cite{Zhao} is nonorthogonal (this being important for description of systems with different coordination numbers \cite{Menon2}) and free of Hubbard-like terms, thus avoiding the self-consistent calculation of occupation numbers and reducing the computation time. This model gives the geometrical and energetical characteristics of small $\mathrm{C_{n}H_{m}}$ and
$\mathrm{C_{n}H_{m}O_{l}}$ molecules, fullerenes, etc. which in general agree with experimental data and DFT calculations. It can be used for simulations of, e. g., large organic molecules, clusters, and bulk carbon-based materials in the cases that \textit{ab initio} calculations are limited by abilities of modern computers.

A serious drawback of the model \cite{Zhao} is, however, unsatisfactory description of C-C bonding in systems with low coordination numbers $K_{c}$ of carbon atoms. For example, the calculated binding energy of the dimer $\mathrm{C_{2}}$ is almost a factor of two greater than its experimental value, while the bond length C-C is much shorter. For the chains $\mathrm{C_{3}}$ and $\mathrm{C_{4}}$, the discrepancy between the theory and experiment is not so large but still rather appreciable (the binding energies are greater than the experimental values by $\approx$ 10\% and $\approx$ 20\% respectively). Thus, the model \cite{Zhao} overestimates the bond strength for carbon atoms with $K_{c}$ = 1 and/or 2 in the absence of adjacent hydrogen and/or oxygen atoms. As a consequence, this model gives qualitatively incorrect results in the cases when accurate values of binding energies of small carbon clusters are needed (e. g., the fragmentation of
fullerene $\mathrm{C_{60}}$ through the loss of a $\mathrm{C_{2}}$ dimer
\cite{Openov2}) or for the systems with large relative number of carbon atoms having  $K_{c}$ = 1 and 2. For example, the 100-atomic graphene fragment appears to be unstable and loses its overall hexagonal structure upon separation of boundary atoms during relaxation (this does not occur if the model \cite{Xu} developed for carbon systems is used). A possible reason for the mentioned shortcomings of the model \cite{Zhao} is that the parameters of the tight-binding Hamiltonian were fitted to DFT calculations and not to the experiment.

The present paper is aimed at the search for such a set of parameters of the nonorthogonal tight-binding model that does not result in contradiction with experimental data for the $\mathrm{C_{2}}$ dimer and a number of other carbon systems, while the energetical and structural characteristics of various $\mathrm{C_{n}H_{m}}$ molecules and clusters agree with experiment no worse or even better than in Ref. \cite{Zhao}. We fit the model parameters based on the criterion of the best correspondence between the calculated and experimental (not DFT-derived) values of bond lengths and binding energies of several selected small $\mathrm{C_{n}H_{m}}$ molecules. The resulting tight-binding potential appears to work well for other, relatively large $\mathrm{C_{n}H_{m}}$ molecules and clusters as well. It also correctly describes the crystalline carbon structures. In this work, we restrict ourselves to hydrocarbons. Generalization to the C-H-O systems will be done later.

The paper is organized as follows. In Section 2, we recall the nonorthogonal tight-binding model for hydrocarbons, describe the alrorithm used to find its parameters, and present the results obtained. In Section 3, we compare the calculated bond lengths and binding energies of various $\mathrm{C_{n}H_{m}}$ molecules and carbon clusters with the corresponding experimental values and the results of the work \cite{Zhao}. In Section 4 we consider bulk carbon structures, diamond and graphene, as well as hydrogen interstitial defects in diamond. Section 5 concludes the paper.

\vskip 10mm

\centerline{\bf 2. Parameters of nonorthogonal tight-binding model for hydrocarbons}

In the tight-binding model, the total energy $E$ for a given set of atomic positions $\left\{\mathbf{R}_{i}\right\}$ is
\begin{equation}
E=E_{\mathrm{el}}+E_{\mathrm{rep}},
\end{equation}
where
\begin{equation}
E_{\mathrm{el}}={\displaystyle \sum_{n,\sigma (\mathrm{occ})}}\varepsilon_{n}
\end{equation}
is the quantum-mechanical electronic ({}``band'') component of $E$ which is the sum of one-electron energies $\varepsilon_{n}$ for the occupied states (in the absence of magnetic field, $\varepsilon_{n}$ does not depend on the spin projection $\sigma =$ $\uparrow$ or $\downarrow$),
\begin{equation}
E_{\mathrm{rep}}={\displaystyle \sum_{i}}{\displaystyle \sum_{j>i}}\phi(R_{ij})
\end{equation}
is the classical component of $E$, being equal to the sum of pairwise ionic repulsive potentials $\phi(R_{ij})=\phi(\left|\mathbf{R}_{i}-\mathbf{R}_{j}\right|)$.

The energy spectrum $\left\{ \varepsilon_{n}\right\}$ is found from the solution of the stationary Schr\"{o}dinger equation
\begin{equation}
\hat{H}\Psi_{n}(\mathbf{r})=\varepsilon_{n}\Psi_{n}(\mathbf{r})\end{equation}
by the expansion of eigenfunctions
\begin{equation}
\Psi_{n}(\mathbf{r})={\displaystyle
\sum_{i,\alpha}}C_{i\alpha}^{n}\varphi_{i\alpha}(\mathbf{r})\end{equation}
in terms of the nonorthogonal atomic orbitals $\left\{ \varphi_{i\alpha}(\mathbf{r})\right\} $, where $i$ is the atomic number, $\alpha$ labels the type of atomic orbital
($1S$ orbitals of hydrogen atoms and $2S,\,2P_{x},\,2P_{y},\,2P_{z}\,$ orbitals of carbon atoms are taken into account). Upon substitution of Eq. (5) into Eq. (4), multiplication
by $\varphi_{j\beta}^{*}(\mathbf{r})$ from the left side and integration over
$\mathbf{r}$, equation for the one-electron self-energies takes the form
\begin{equation}
{\displaystyle \sum_{i,\alpha}}(H_{i\alpha}^{j\beta}-\varepsilon_{n}S_{i\alpha}^{j\beta})C_{i\alpha}^{n}=0 ,
\end{equation}
where
\begin{equation}
H_{i\alpha}^{j\beta}=\int d\mathbf{r}\varphi_{j\beta}^{*}(\mathbf{r})\hat{H}\varphi_{i\alpha}(\mathbf{r})
\end{equation}
are the matrix elements of the Hamiltonian,
\begin{equation}
S_{i\alpha}^{j\beta}=\int d\mathbf{r}\varphi_{j\beta}^{*}(\mathbf{r})\varphi_{i\alpha}(\mathbf{r})
\end{equation}
are overlap integrals for atomic orbitals. The one-electron Schr\"{o}dinger equation
(4) is thus reduced to the generalized eigenvalue problem (6) which is solved numerically. The number of equations in the system (6) equals to the total number of atomic orbitals involved, i. e., $N_{\mathrm{H}}+4N_{\mathrm{C}}$, where $N_{\mathrm{H}}$ and $N_{\mathrm{C}}$ is the number of hydrogen and carbon atoms respectively. Molecular orbitals (5) are occupied (according to the Fermi-Dirac distribution function and Pauli principle) by electrons, the number of which is $N_{\mathrm{H}}+4N_{\mathrm{C}}$ as well. Note that in the orthogonal tight-binding models, matrix $S_{i\alpha}^{j\beta}$ is diagonal in $(i,j)$ and $(\alpha,\beta)$.

If the nonorthogonal tight-binding model is used in molecular dynamics simulations, the force $\mathbf{F}_{k}=-\partial E/\partial\mathbf{R}_{k}$ acting on the $k$-th atom is found after calculation of eigenenergies $\varepsilon_{n}$ and eigenvectors $C_{i\alpha}^{n}$. As follows from Eq. (6),
\begin{equation}
\frac{\partial\varepsilon_{n}}{\partial\mathbf{R}_{k}}=\frac{{\displaystyle
\sum_{i,\alpha}}{\displaystyle
\sum_{j,\beta}}C_{j\beta}^{n*}(\partial
H_{i\alpha}^{j\beta}/\partial\mathbf{R}_{k}-\varepsilon_{n}\partial
S_{i\alpha}^{j\beta}/\partial\mathbf{R}_{k})C_{i\alpha}^{n}}{{\displaystyle
\sum_{i,\alpha}}{\displaystyle
\sum_{j,\beta}}C_{j\beta}^{n*}S_{i\alpha}^{j\beta}C_{i\alpha}^{n}}.
\end{equation}
Explicit expressions for $S_{i\alpha}^{j\beta}$ and $H_{i\alpha}^{j\beta}$ (see below) greatly simplify simulations of relatively large systems.

For $H_{i\alpha}^{j\beta}$ the following parametrization is used \cite{Hoffmann}:
\begin{equation}
H_{i\alpha}^{j\beta}=\frac{1}{2}K_{ij}S_{i\alpha}^{j\beta}(H_{\alpha}+H_{\beta}) ,
\end{equation}
where
\begin{equation}
K_{ij}=\left\{ \begin{array}{l} 1, \, i=j\\
K_{ij}^{0}\exp\left[-\delta_{ij}\left(R_{ij}-R_{ij}^{0}\right)\right],\,
i\neq j\end{array}\right.
\end{equation}
(note that $H_{i\alpha}^{i\beta}=H_{\alpha}$ at $\alpha=\beta$ and
$H_{i\alpha}^{i\beta}=0$ at $\alpha\neq\beta$ since $S_{i\alpha}^{i\beta}=\delta_{\alpha\beta}$). There are three different parameters $H_{\alpha}$, according to the number of different atomic orbitals $\alpha=1S,\,2S,\,2P$, and three different parameters
$K_{ij}^{0},\,\delta_{ij},\, R_{ij}^{0}$ each, according to the number of pairs of atoms of different sorts $ij$ = HH, HC, CC (note that in Ref. \cite{Zhao}, the coefficients $\delta_{ij}$ were taken to be the same for all pairs of atoms). The values of $S_{i\alpha}^{j\beta}$ are obtained analytically \cite{Roothaan,Slater} using the expressions for Slater orbitals overlaps:
\begin{equation}
\varphi_{1S}(\mathbf{r})=\sqrt{\frac{\xi_{1S}^{3}}{\pi}}\exp\left(-\xi_{1S}r\right),\end{equation}
\begin{equation}
\varphi_{2S}(\mathbf{r})=\sqrt{\frac{\xi_{2S}^{5}}{3\pi}}\cdot
r\cdot\exp\left(-\xi_{2S}r\right),\end{equation}
\begin{equation}
\varphi_{2P_{\gamma}}(\mathbf{r})=\sqrt{\frac{\xi_{2P}^{5}}{\pi}}\cdot\gamma\cdot\exp\left(-\xi_{2P}r\right),\end{equation}
where $\gamma=x,y,z$.

Pair potentials in Eq. (3) for $E_{\mathrm{rep}}$ are taken in the form
\begin{equation}
\phi\left(R_{ij}\right)=\phi_{ij}^{0}\exp\left[-\beta_{ij}\left(R_{ij}-R_{ij}^{0}\right)\right]\end{equation}
(there are three parameters $\phi_{ij}^{0}$ and $\beta_{ij}$ each). So, the total number of fitting parameters ($H_{\alpha}$, $K_{ij}^{0},\,\delta_{ij},\, R_{ij}^{0}$, $\phi_{ij}^{0}$, $\beta_{ij}$, $\xi_{1S}$, $\xi_{2S}$, $\xi_{2P}$) is 21. We remain unchanged (see Ref. \cite{Zhao}) the parameters $H_{1S}=-10.70$ eV, $K_{\mathrm{HH}}^{0}=1.68$ eV, $\delta_{\mathrm{HH}}=0.13$ \AA $^{-1}$, $R_{\mathrm{HH}}^{0}=0.75$ \AA , $\phi_{\mathrm{HH}}^{0}=0.78$ eV,
$\beta_{\mathrm{HH}}=6.84$ \AA $^{-1}$, and $\xi_{1S}=2.456644$ \AA $^{-1}$
describing the H-H interaction. As mentioned above, our main purpose was to considerably improve the correspondence between theory and experiment for small carbon clusters and several other purely carbon systems. We were unable to do it through changes in the parameters of C-C interactions only, because of drastic decrease in accuracy of the results obtained for various hydrocarbon molecules $\mathrm{C_{n}H_{m}}$. This is why we changed the parameters of C-H interactions along with the parameters of C-C interactions. Search for new values of those parameters was based on the criterion of the best correspondence between the calculated and experimental interatomic distances, binding energies, and some vibration frequencies of the following clusters, molecules, and radicals: $\mathrm{C_{2}}$, $\mathrm{C_{3}}$, $\mathrm{C_{4}}$, $\mathrm{CH}$, $\mathrm{CH_{2}}$, $\mathrm{CH_{3}}$, $\mathrm{CH_{4}}$, $\mathrm{C_{2}H_{2}}$, $\mathrm{C_{6}H_{6}}$.

Since the number of fitting parameters exceeded the total number of model parameters,
the latter were derived by minimization of the function $\Phi(H_{2S},H_{2P},K_\mathrm{{CC}}^{0},K_\mathrm{{CH}}^{0}, ...)$ which we chose as a quadratic form of the differences between the corresponding calculated and experimental values. Usual minimization techniques such as the gradient or Newton methods can hardly be applied to the function $\Phi$ because 1) this function has many local minima and 2) cusps or even discontinuities in the dependences of the physical quantities on the model parameters are possible, thus making the use of those methods questionable since they imply that the function $\Phi$ is sufficiently smooth. For this reason, in order to find the global minimum of the function $\Phi$ we used the simplest version of the Monte-Carlo method. Our strategy was to explore the neighbourhood of the last {}``best'' (i. e., having the minimum value of $\Phi$) point in the parameter space among the points considered up to a given iteration step. The coordinates of each new point were determined using the generator of pseudo-random numbers. The size of the region for the search of new points was periodically changed, which is necessary both for refining the coordinates of the $\Phi$ minimum found at the preceding step and for finding other, probably more deep minima.

Finally we obtained the following values of the model parameters: $H_{2S}=-16.157972$ eV, $H_{2P}=-10.078261$ eV, $K_{\mathrm{CC}}^{0}=2.060290$ eV,
$K_{\mathrm{CH}}^{0}=1.763801$ eV, $\delta_{\mathrm{CC}}=0.164262$ \AA $^{-1}$, $\delta_{\mathrm{CH}}=0.014350$ \AA $^{-1}$, $R_{\mathrm{CC}}^{0}=1.582565$ \AA , $R_{\mathrm{CH}}^{0}=1.045120$ \AA , $\phi_{\mathrm{CC}}^{0}=0.943505$ eV, $\phi_{\mathrm{CH}}^{0}=0.561102$ eV, $\beta_{\mathrm{CC}}=4.912617$ \AA $^{-1}$, $\beta_{\mathrm{CH}}=9.433587$ \AA $^{-1}$, $\xi_{2S}=2.991164$ \AA $^{-1}$, $\xi_{2P}=3.857861$ \AA $^{-1}$. They do not differ much from those given in Ref. \cite{Zhao}, with the exception of $\delta_{\mathrm{CH}}$ which is a factor of $\approx$ 9 smaller that in Ref. \cite{Zhao}. We draw attention to the fact that $\xi_{2S}<\xi_{2P}$, while $\xi_{2S}>\xi_{2P}$ in Ref. \cite{Zhao}.

\vskip 10mm

\centerline{\bf 3. Binding energies and structures of C$_n$H$_m$ molecules}

The binding energies $E_{b}$ of clusters and molecules $\mathrm{C_{n}H_{m}}$
were determined as
\begin{equation}
E_{b}\left(\mathrm{C}_{n}\mathrm{H}_{m}\right)=nE(\mathrm{C})+mE(\mathrm{H})-E(\mathrm{C}_{n}\mathrm{H}_{m})~,
\end{equation}
where $E\left(\mathrm{C}_{n}\mathrm{H}_{m}\right)$ is the total energy of the system, $E\left(\mathrm{C}\right)$ and $E\left(\mathrm{H}\right)$ are the energies of isolated carbon and hydrogen atoms, respectively. In the tight-binding model, they are $E(\mathrm{C})=2E_{2S}+2E_{2P}$ and $E(\mathrm{H})=E_{1S}$. The specific (per atom) value of $E_{b}$ is found from Eq. (16) through deviding by the total number of atoms in the system, $(n+m)$. The results obtained for several clusters and molecules are listed in Table 1 along with the corresponding experimental values and the values calculated using the set of the model parameters from Ref. \cite{Zhao}. Note that theoretical values of $E_{b}$ are given without account for the zero-point energy which is, as a rule, about $\sim0.1$ eV/atom and results in a small decrease of $E_{b}$.

From Table 1 one can see that the binding energies for the dimer $\mathrm{C_{2}}$, the trimer $\mathrm{C_{3}}$, and the chains $\mathrm{C_{4}}$ and $\mathrm{C_{5}}$ are much closer to the experimental values than those calculated with the model parameters from Ref. \cite{Zhao} (for the dimer $\mathrm{C_{2}}$ there is a qualitative change in the value of $E_{b}$ almost by a factor of two as compared with Ref. \cite{Zhao}). For the majority of small $\mathrm{C_{n}H_{m}}$ molecules and $\mathrm{C_{8}H_{8}}$ cluster our values of $E_{b}$ are also closer to experimental ones than those calculated with the parameters of Ref. \cite{Zhao}, see Table 1. This is true for relatively large molecules
$\mathrm{C_{n}H_{m}}$ as well. For example, we obtained $E_{b}=$ 5.09; 4.31; 5.03 eV for $\mathrm{C_{10}H_{8}}$ (naphthalene), $\mathrm{C_{10}H_{16}}$ (adamantane), and $\mathrm{C_{12}H_{10}}$ (acenaphthene) respectively, while the experimental and calculated with the parameters from Ref. \cite{Zhao} values are, respectively, $E_{b}=$ 5.07; 4.35; 5.03 eV \cite{Experiment1} and $E_{b}=$ 5.17; 4.40; 5.09 eV.

Table 1 shows also theoretical and experimental values of interatomic distances C-C and C-H. One can see that for the dimer $\mathrm{C_{2}}$ there is much better correspondence with the experiment as compared with Ref. \cite{Zhao}. For the molecules and large clusters $\mathrm{C_{n}H_{m}}$, the deviations from the experimental values, as a rule, do not exceed several hundredth of \AA , such an accuracy being sufficient in simulations of a broad range of hydrocarbons. However, it should be pointed out that the C-H bond lengths calculated in Ref. \cite{Zhao} are often somewhat closer to the experimental ones. As mentioned above, this is at a price of unsatisfactory description of C-C interactions in some carbon systems, while our model adequately describes, e.g., the two-dimensional graphene fragments which preserve their overall hexagonal structure upon relaxation, being just slightly distorted.

We have calculated the binding energies and bond lengths in several fullerenes. For the smallest possible fullerene $\mathrm{C_{20}}$ \cite{Prinzbach}, we obtained $E_{b}=$ 6.31 eV/atom, in agreement with the values of $E_{b}=6.08$ eV/atom \cite{Davydov} and $E_{b}=6.36$ eV/atom \cite{Jones} found, respectively, within the orthogonal tight-binding model \cite{Xu} and by DFT with gradient corrections (we are not aware of experimental $E_{b}$ value for the fullerene $\mathrm{C_{20}}$). The calculated minimal and maximal C-C bond lengths, $l_{min}=$ 1.44 \AA~  and $l_{max}=$ 1.52 \AA~ coincide with those obtained within the orthogonal tight-binding model \cite{Xu} and agree well with recent {\it ab initio} calculations for various exchange-correlation potentials \cite{An} ($l_{min} = 1.40 - 1.43$  \AA, $l_{max} = 1.51 - 1.52$ \AA ). Note that in the model \cite{Zhao}, the values of $E_{b}=$ 5.89 eV/atom, $l_{min}=1.45$ \AA   , and $l_{max}=1.61$ \AA ~ differ from the results of Refs. \cite{Davydov,Jones,An} much greater.

As for the different isomers of $\mathrm{C_{20}}$, we have found that the bowl isomer is energetically more favourable than the cage, its binding energy being $E_{b}=$ 6.55 eV/atom, while the ring isomer with $E_{b}=$ 6.81 eV/atom is more stable than the bowl one. In the absence of experimental information on the relative stability of $\mathrm{C_{20}}$ isomers, various theoretical approaches give conflicting results. However several authors arrived at the same sequence of $\mathrm{C_{20}}$ isomers as described above, see references in Ref. \cite{Sokolova}.

Next we calculated the coagulation energy $\bigtriangleup E=2E[$C$_{20}]-E[($C$_{20})_{2}]$ of two fullerenes C$_{20}$ in the {\it open}-[2+2] isomer of a cluster molecule (C$_{20})_{2}$ \cite{Choi}. We found $\bigtriangleup E=4.9$ eV. This value coincides with that obtained in the orthogonal tight-binding model \cite{Xu,Openov3} and does not
differ much from the value of $\bigtriangleup E=6.3$ eV obtained in the DFT \cite{Choi}. The intercluster bond length $l=1.37$ \AA~  agrees well with both orthogonal tight-binding model \cite{Xu} ($l=1.35$ \AA~ \cite{Podlivaev2}) and DFT ($l=1.34$ \AA~ \cite{Choi}).

For fullerene $\mathrm{C_{60}}$ \cite{Kroto}, the binding energy and the bond lengths are $E_{b}=$ 7.01 eV/atom and $l=1.41, 1.48$ \AA , respectively, in good agreement with the experimental values of $E_{b} =$ 6.93 eV/atom and $l=1.40, 1.46$ \AA~ \cite{Experiment1}. The model \cite{Zhao} gives $E_{b}=6.71$ eV/atom and $l=1.42, 1.53$ \AA , in a poorer correspondence with the experiment. Meanwhile, in the model \cite{Zhao}, the HOMO-LUMO gap $\bigtriangleup=1.91$ eV is closer to the experimental value $\bigtriangleup= 1.6 - 1.8$ eV \cite{Dresselhaus} than our result $\bigtriangleup=1.15$ eV. So, the suggested set of parameters of the nonorthogonal tight-binding model is more suited to calculations of binding energies and bond lengths than electron characteristics.

For fullerene $\mathrm{C_{70}}$ we obtained $E_{b}=$ 7.04 eV/atom and $l=1.41 - 1.49$ \AA , again in much better agreement with the experimental data $E_{b}=$ 6.97 eV/atom \cite{Kiyobayashi} and $l=1.37 - 1.48$ \AA~ \cite{Roth} than the results of the model \cite{Zhao} $E_{b}=$ 6.73 eV/atom and $l=1.41 - 1.55$ \AA~ (we are not aware of reliable experimental data on the heats of formation and bond lengths in fullerenes other than $\mathrm{C_{60}}$ and $\mathrm{C_{70}}$).

\vskip 10mm

\centerline{\bf 4. Bulk crystalline forms of carbon}

We have also calculated the binding energies and bond lengths for two crystalline forms of carbon, diamond and graphene. We made use of periodic boundary conditions and performed scaling with respect to dimensions of a supercell. For the binding energy and bond length in diamond we obtained $E_{b}=$ 7.36 eV/atom and $l = 1.54$ \AA , in excellent agreement with experimental values $E_{b}=$ 7.35 eV/atom and $l = 1.54$ \AA . On the other hand, the model \cite{Zhao} gives $E_{b}=$ 6.58 eV/atom and $l = 1.61$ \AA . These results differ substantially from experimental ones.

For graphene our tight-binding potential results in $E_{b}=$ 7.36 eV/atom and $l = 1.45$ \AA . Since the experimental value of $E_{b}$ in graphite is 7.37 eV/atom and the weak interlayer coupling is abouth 0.02 eV/atom, for graphene one has $E_{b}=$ 7.35 eV/atom, in close correspondence with our value of $E_{b}$. Although the calculated bond length is somewhat large than the experimental one, $l = 1.42$ \AA , we note that the model \cite{Zhao} leads to much stronger deviations from the experiment in both $E_{b}=$ 7.06 eV/atom and $l = 1.48$ \AA . It is particularly remarkable that a rather good description of energetics and structure of {\it bulk phases} of carbon was given within the model whose parameters had been derived based on a comparison with experimental data for {\it small clusters and molecules}. This points to a good transferability of our tight-binding model, at least in what concernes pure carbon systems.

Now we turn to a much more subtle problem, the hydrogen interstitials in diamond \cite{Goss}. Unlike Si and Ge, there is no experimental data on isolated interstitial hydrogen in diamond. Based on the experiments with muonium, it is generally believed that there are at least two stable interstitial sites, tetrahedral (T) and bond-centred (BC), the latter being 1 - 2 eV lower in energy (see Fig. 1 in Ref, \cite{Goss}).

We made calculations for 64- and 216-atom supercells. The results vary only slightly with the size of a supercell. We have found that the BC-interstitial corresponds to a minimum of the total energy as a function of atomic coordinates, i.e., all oscillation frequencies are real.
The formation energy of this interstitial is $E_{f}=$ 1.4 eV. Two C-H bond lengths are both 1.10 \AA , in accordance with calculations of other authors, 1.05 - 1.17 \AA ~ \cite{Goss}. We have also identified several other stable interstitial sites, including the H-interstitial with the energy 1.3 eV higher than that of the BC-interstitial (to be compared with the values 1.5 - 1.9 eV obtained within DFT \cite{Goss}).

Detailed description of our results on hydrogen interstitials in diamond goes beyond the scope of this paper. We note, however, that the T-interstitial appeared to be the saddle point lying 1.4 eV above the BC-interstitial. Upon relaxation, the hydrogen atom moved to one of two stable sites close to the C-site \cite{Goss}. The reasons why the T-interstitial appeared to be not a local minimum of the total energy are not quite clear to us. The only essential difference between the multi-coordinated hydrogen atoms in H- and T-sites is that in the latter case the bonds C-H do not lie in the same plane.

\vskip 10mm

\centerline{\bf 5. Conclusions}

The set of parameters for the Hamiltonian of the nonorthogonal tight-binding model derived in this work not only provides rather accurate description of C-H interactions in hydrocarbons C$_n$H$_m$, but also allows one to simulate the structure and energetics of both small (C$_{2}$, C$_{3}$, C$_{4}$) and relatively large (fullerenes C$_{20}$ and C$_{60}$, graphene fragments) carbon clusters, as well as bulk crystalline forms of carbon (diamond, graphene). The accuracy of the method is sufficient for many purposes. Analytical dependences of the overlap integrals on atomic coordinates greatly simplify the calculation of forces acting on the atoms. This makes possible the molecular dynamics simulations of relatively large systems for which {\it ab initio} calculations are
problematic due to the limited computer power. Besides, the evolution of small clusters can be followed on a microsecond time scale \cite{Openov4, Maslov} (the characteristic times are about 1 - 10 ps for {\it ab initio} molecular dinamics). On the other hand, this tight-binding potential fails to describe
the tetrahedral hydrogen interstitial in diamond, although the bond-centred and hexagonal interstitials are well reproduced. Hence it may be not applicable to the systems with multi-coordinated hydrogen atoms where the C-H bonds do not lie in the same plane. There are, however, just few examples of such systems. Later we plan to generalize our approach to the C-H-O systems.

\newpage

Table 1. Binding energies $E_b$ and bond lengths $l_{\mathrm{CC}}$, $l_{\mathrm{CH}}$ for some C$_n$H$_m$ molecules.

\vskip 5mm

$^*$ Experimental values of $E_{b}\left[\mathrm{eV/atom}\right]$ are obtained from known heats of formation
$\Delta_{f}H^{\circ}\left[\mathrm{kJ/mol}\right]$ \cite{Experiment1} making use of relations
$E_{b}=\frac{1}{(n+m)}\cdot\left\{nE_{\mathrm{C}}^{0}+mE_{\mathrm{H}}^{0}-1.0364\cdot10^{-2}\cdot\Delta_{f}H^{\circ}\right\}$, where $E_{\mathrm{C}}^{0}=7.3768$ eV, $E_{\mathrm{H}}^{0}=2.375$ eV \cite{Brenner}.

\begin{tabular}{|c|c|c|c|c|c|c|}
\hline
Formula &  \multicolumn{3}{|c|}{Binding energy $E_{b}$, eV/atom} &  \multicolumn{3}{|c|}{Bond lengths, {\AA}} \\  \cline{2-7}
and & Potential & Present & Experimental & Potential & Present & Experimental  \\
Name &  from \cite{Zhao} & work & data$^{*}$ \cite{Experiment1,Experiment2} &  from \cite{Zhao} & work & data \cite{Experiment2} \\        \hline

C$_{2}$ & & & & & & \\
Carbon & $5.78$ & $3.15$ & $3.12$ & $1.157$ & $1.230$ & $1.243$ \\
dimer & & & & & & \\
\hline

C$_{3}$ & & & & & & \\
Carbon & $5.10$ & $4.72$ & $4.54$ & $1.300$ & $1.301$ & $1.277$ \\
trimer & & & & & & \\
\hline

C$_{4}$ & & & & & & \\
Carbon & $5.86$ & $5.09$ & $4.88$ & $1.187$ & $1.296$ & --- \\
chain  &        &        &        & $1.513$ & $1.354$ &  \\
\hline

C$_{5}$ & & & & & & \\
Carbon & $6.07$ & $5.68$ & $5.35$ & $1.263$ & $1.273$ & --- \\
chain  &        &        &        & $1.338$ & $1.348$ &  \\
\hline

CH & & & & & & \\
Methylidyne & $1.90$ & $1.87$ & $1.80$ & $1.089$ & $1.081$ & $1.120$ \\
 & & & & & & \\
\hline

CH$_{2}$ & & & & & & \\
Methylene & $2.80$ & $2.75$ & $2.69$ & $1.079$ & $1.080$ & $1.085$ \\
& & & & & & \\
\hline

CH$_{4}$ & & & & & & \\
Methane & $3.66$ & $3.40$ & $3.51$ & $1.089$ & $1.100$ & $1.094$ \\
& & & & & & \\
\hline

C$_{2}$H$_{2}$ & & & & & & \\
Acetylene & $4.87$ & $4.54$ & $4.29$ & C-C $1.199$ & C-C $1.226$ & C-C $1.203$ \\
 & & & &                                                C-H $1.064$ & C-H $1.079$ & C-H $1.063$ \\
\hline

C$_{2}$H$_{4}$ & & & & & & \\
Ethylene & $4.19$ & $3.96$ & $3.94$ & C-C $1.315$ & C-C $1.327$ & C-C $1.339$ \\
 & & & &                                              C-H $1.091$ & C-H $1.097$ & C-H $1.086$ \\
\hline

C$_{3}$H$_{4}$ & & & & & & \\
Allene & $4.56$ & $4.32$ & $4.22$ & C-C $1.310$ & C-C $1.323$ & C-C $1.308$ \\
& & & &                                           C-H $1.096$ & C-H $1.100$ & C-H $1.087$ \\
\hline

C$_{6}$H$_{6}$ & & & & & & \\
Benzene & $4.95$ & $4.82$ & $4.79$ & C-C $1.422$ & C-C $1.407$ & C-C $1.397$ \\
 & & & &                                              C-H $1.092$ & C-H $1.095$ & C-H $1.084$ \\
\hline

C$_{8}$H$_{8}$ & & & & & & \\
Cubane & $4.37$ & $4.42$ & $4.47$ & C-C $1.645$ & C-C $1.570$ & C-C $1.571$ \\
 & & & &                                            C-H $1.075$ & C-H $1.082$ & C-H $1.097$ \\
\hline

\end{tabular}

\end{document}